\documentstyle[prl,aps,multicol,psfig,epsfig]{revtex}

\begin{document}
\draft
\begin{multicols}{2}
{\bf Pudalov {\em et al}. Reply}. In the recent Letter \cite{gm},
we reported {\em direct} measurements of the renormalized spin
susceptibility $\chi^*$ in Si-MOS samples, and showed that
$\chi^*$ increases gradually  as the electron density $n$
decreases. It remains finite at the critical density $n_c$ of the
apparent metal-insulator transition in 2D (2D MIT).  In the
preceding Comment\cite{KSDcomment}, Kravchenko, Shashkin, and
Dolgopolov (KSD) claim that our results: (i) are consistent with
their {\em indirect} data \cite{SKDprl}, and (ii) support their
idea that $\chi^*$ tends to diverge at a ``sample-independent
electron density $n_\chi$, which coincides with the  critical
density $n_c$" for the 2D MIT in the sample studied in
Ref.~\cite{SKDprl}.  We disagree with their claims as explained
below.

The manner in which the three sets of data are plotted in Fig.~1
of Ref.~\cite{KSDcomment} obscures the important systematic
difference in the density dependences of $\mu_BB_c\equiv \pi
\hbar^2 n \mu_B/(g^*m^*)$,   where $g^*m^*\propto \chi^*$
\cite{gm}. Our data alone are replotted in Fig.~1 in the same
units. The inset shows that $B_c$ depends  on $n$ almost linearly
for $n>5\times10^{11}$cm$^{-2}$. At lower densities, $n \leq$
2$\times10^{11}$cm$^{-2}$, there are clear deviations from the
KSD-conjecture $B_c \propto (n - n _{\chi})$ \cite{KSDcomment},
depicted as the dash-dotted line.  Our data remain {\em finite} at
the density $n_{\chi}=8\times 10^{11}$cm$^{-2}$ where $\chi^*$ is
thought \cite{KSDcomment} to diverge.  The upper estimate
$g^*m^*/2m_b \approx 7$ at $n=0.77\times 10^{11}$cm$^{-2}$ can be
obtained from the phase of SdH oscillations for sample Si5; for
bigger $g^*m^*$ values, the phase would change by $\pi$ in
contrast to our observations \cite{polariz}.  Thus, we find a
significant difference between our data and that of
Ref.~\cite{KSDcomment}. Clearly, the search for possible critical
behavior of a nonlinear function $1/\chi^*(n)$ requires more
careful consideration, for even the critical range of $n$ is
unknown. Such analysis has been performed by us in
Ref.~\cite{polariz}: we concluded that the divergency of $\chi^*$
is unlikely at $n>0.5\times 10^{11}$cm$^{-2}$.

By extrapolating  $1/\chi^* \rightarrow 0$, KSD made a conclusion
of a spontaneous complete spin polarization of mobile electrons
(the "ferromagnetic instability") at
$n=0.8\times10^{11}$cm$^{-2}$. The absence of any traces of such
instability in our SdH data at $n=n_c$
($=1\times10^{11}$cm$^{-2}$ for Si6-14) was attributed by KSD to
a stronger disorder in  our samples. However, the SdH data for
sample Si5, which is less disordered than any samples studied by
KSD, clearly demonstrate the absence of a ferromagnetic
transition at $n=0.77\times 10^{101}$cm$^{-2}$ and allow to
estimate the spontaneous component of polarization at this $n$ to
be less than 15\% (Fig.~3 of Ref.~\cite{polariz}).

The difference between our results and that of KSD might be due to
 the following reasons:
in Ref.~\cite{gm},  $\chi^*(n)$ is determined from Shubnikov-de
Haas (SdH) oscillations in \underline{weak} crossed magnetic
fields from the difference in the numbers of spin-up and
spin-down electrons. This approach based solely on Landau
quantization provides {\em direct} results, which hold for
arbitrary interaction strength.  In contrast, the data of
Ref.~\cite{SKDprl} are indirect and based on a conjecture that
the magnetoresistance (MR) in \underline{strong} in-plane fields
($g\mu_BB_\parallel \lesssim E_F$) scales as $1/\chi^*$. We have
shown \cite{MR} that the MR, in contrast to the SdH data, depends
not only on $n$, but also on the history-dependent disorder in a
sample, and differs from sample to sample. The effect of disorder
on the MR becomes especially strong at low densities and high
resistivities $\rho \sim h/e^2$. Thus, attributing
$R(B_\parallel)$ solely to the spin polarization of mobile
electrons is dangerous at best.

The KSD concern  about applicability of the Lifshits-Kosevitch
(LK) formula to strongly interacting systems has been addressed in
Refs.~\cite{gorkov}. It was shown that the LK formula with
renormalized $g^*$ and $m^*$ holds for an arbitrary interaction
strength provided the system remains Fermi-liquid and the
amplitude of oscillations is small.

\begin{figure}
\centerline{\psfig{figure=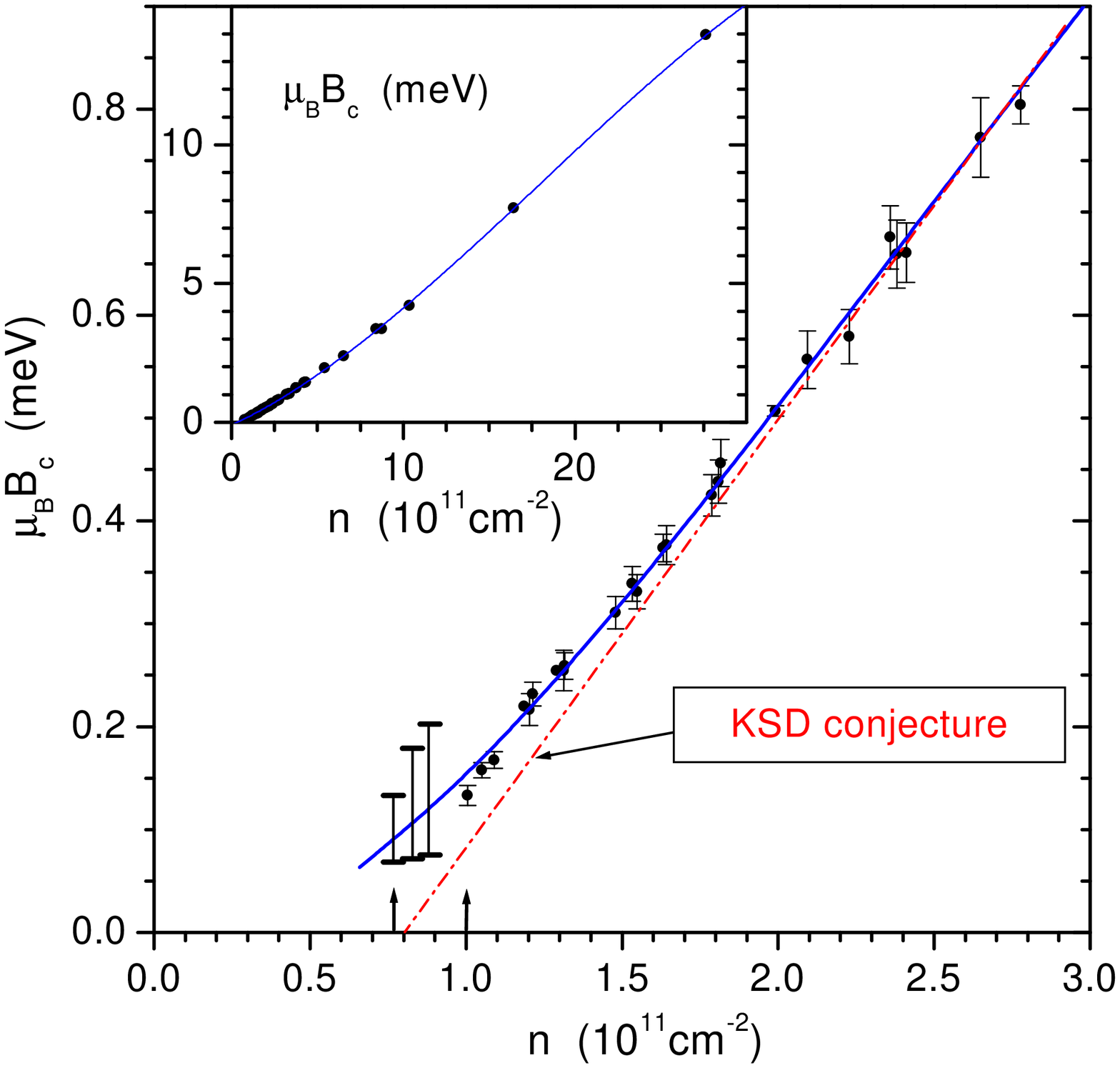,width=230pt,height=210pt}}
\vspace{0.1in}
\begin{minipage}{3.2in}
\caption{
$\mu_B B_c=\pi \hbar^2 n \mu_B/(g^*m^*)=0.63n(2m_b/g^*m^*)$
meV$10^{-11}$cm$^{2}$  plotted
vs $n$ using direct $g^*m^*(n)$-data \protect\cite{gm,polariz}.
Solid line is a guide to the eye. Vertical arrows depict $n=n_c$
for samples Si6-14 and Si5. Dash-dotted line represents the KSD
conjecture \protect\cite{KSDcomment}.}
\end{minipage}
\label{Fig.1}
\end{figure}
\vspace{-0.05in}

{\em To summarize}  (i) there is a systematic disagreement at low
densities between our {\em direct} results on $\chi^*$ and {\em
indirect} data by KSD; (ii) both $\chi^*$ and the period of SdH
oscillations show absence of a complete spin polarization down to
$n=0.77\times 10^{11}$cm$^{-2}$; this rules out the spontaneous
ferromagnetic transition  at $n=0.8\times10^{11}$cm$^{-2}$
suggested in Refs.~\cite{KSDcomment,SKDprl}.

Authors acknowledge support by NSF, FWF Austria, INTAS, NATO, and
RFBR.

 V.\ M.\ Pudalov, M.\ Gershenson, H.\ Kojima, N.\ Busch,
 E.\ M.\ Dizhur, G.\ Brunthaler, A.\
Prinz, and G.\ Bauer.

\end{multicols}
\end{document}